\documentclass[manuscript, nonacm]{acmart}

\usepackage[utf8]{inputenc}
\usepackage{graphicx, color}
\newcommand{\ket}[1]{|#1\rangle}

\usepackage[linesnumbered, ruled, lined, noend]{algorithm2e}
\SetKwFunction{simplify}{simplify}
\SetKwFunction{RepetitionSearch}{repetition\_search}
\SetKwFunction{RepetitionVerify}{repetition\_verify}
\SetKwFunction{lg}{log2}
\SetKwFunction{len}{len}
\SetKwFunction{set}{set}
\SetKwFunction{alist}{list}

\usepackage{hyperref}
\usepackage[caption=false]{subfig}

\setcopyright{none}

\AtBeginDocument{%
  }

\begin{document}

\title{Quantum Multiplexer Simplification for State Preparation}

\author{José A. de Carvalho}
\author{Carlos A. Batista}
\affiliation{
\institution{Centro de Informática, Universidade Federal de Pernambuco}
\city{Recife}
\state{Pernambuco}
\country{Brazil}}
\author{Tiago M.L. de Veras}
\affiliation{
\institution{Departamento de Matemática, Universidade Federal Rural de Pernambuco}
\city{Recife}
\state{Pernambuco}
\country{Brazil}}
\author{Israel F. Araujo}
\affiliation{
\institution{Departamento de Eletrônica e Sistemas, Universidade Federal de Pernambuco}
\city{Recife}
\state{Pernambuco}
\country{Brazil}}
\affiliation{
\institution{Data Cybernetics}
\city{Landsberg am Lech}
\country{Germany}}
\author{Adenilton J. da Silva}
\email{ajsilva@cin.ufpe.br}
\affiliation{
\institution{Centro de Informática, Universidade Federal de Pernambuco}
\city{Recife}
\state{Pernambuco}
\country{Brazil}}

\begin{abstract}
The initialization of quantum states or Quantum State Preparation (QSP) is a basic subroutine in quantum algorithms. In the worst case, general QSP algorithms are expensive due to the application of multi-controlled gates required to build the quantum state. Here, we propose an algorithm that detects whether a given quantum state can be factored into substates, increasing the efficiency of compiling the QSP circuit when we initialize states with some level of unentanglement. The simplification is done by eliminating controls of quantum multiplexers, significantly reducing circuit depth and the number of CNOT gates with a better execution and compilation time than the previous QSP algorithms. Considering efficiency in terms of depth and number of CNOT gates, our method is competitive with the methods in the literature.  However, when it comes to run-time and compilation efficiency, our result is significantly better, and the experiments show that by increasing the number of qubits, the gap between the temporal efficiency of the methods increases.
\end{abstract}




\maketitle

\section{Introduction}

Quantum computing is an emerging and interdisciplinary area that studies tasks that one can solve more efficiently by quantum systems in information processing~\cite{Feynman1982,benioff1982quantum,Nielsen_Chuang_2010}. For instance, quantum algorithms provide solutions to solve some problems more efficiently~\cite{shor1999polynomial, PhysRevLett.127.180501}. The first quantum processors are Noisy Intermediate-Scale Quantum (NISQ) devices because of the number of qubits and noisy operations~\cite{preskill2018quantum}. Over the years, NISQ devices have improved, and the computer science community can contribute to system software development for NISQ and future fault-tolerant quantum devices. The operations available in current gate-based NISQ devices include single-qubit
gates and CNOT gates.

A quantum compiler must decompose any quantum operation into the quantum device instruction set. To initialize a variable in a quantum device, we encounter unexpected challenges that do not exist in the classical counterpart. The no-cloning theorem~\cite{1982Natur.299..802W} shows that it is impossible to copy quantum states, and we must load the information whenever we need to perform a computation. Besides that, the phenomenon of decoherence~\cite{PhysRevLett.77.3240} makes it essential to reload the information periodically. Therefore, a significant cost in the state preparation procedure can compromise quantum speedup.

\subsection{Quantum State Preparation}
Algorithms for quantum state preparation have been studied for more than 20 years~\cite{ventura1999initializing, grover2000synthesis, long2001efficient, Trugenberger2001, mottonen2005transformation, Soklakov2006, 1629135, PhysRevA.83.032302, 10.1109/TCAD.2023.3297972} and are an indispensable subroutine when we need to load classical data into a quantum device. 
A $2^n$ complex vector serves as a literal representation of a $n$-qubit quantum state. Quantum State Preparation algorithms (QSP) produce circuits to initialize a complex vector $\vec{x} =[x_0\ \dots, x_{2^{n}-1}]$ into a quantum variable $\ket{x}=\sum_{i=0}^{{2^n}-1} x_i\ket{i}$, where $x_i$ is the amplitude associated with $\ket{i}$. QSP requires a non-trivial compilation step ~\cite{long2001efficient, grover2000synthesis, mottonen2005transformation, PhysRevA.83.032302,10.1109/TCAD.2023.3297972}, and the resulting quantum circuit requires $O(2^n)$ quantum operations.

Strategies to optimize the compilation of QSP algorithms include a trade-off between circuit depth and circuit size~\cite{araujo_divide_2021, araujo_configurable_2023, gui2024spacetime}, approximated initialization~\cite{10.1109/TCAD.2023.3297972, nakaji2022approximate}, and techniques for sparse quantum states~\cite{gleinig2021efficient, mozafari2022efficient}. One can reduce the executable program size or the circuit depth if the state is not fully entangled. However, this reduction requires a search across the space of quantum bits and the successive use of linear algebra subroutines~\cite{10.1109/TCAD.2023.3297972}. Regarding the compilation strategy, an abstract tree can be used to produce the circuit~\cite{araujo_divide_2021}, a quantum search-based strategy~\cite{grover2000synthesis}, or load amplitudes with an iterative method~\cite{ventura1999initializing, park2019circuit}.

The quantum state preparation proposed in this work uses an abstract syntax tree as an intermediate representation~\cite{araujo_divide_2021, araujo_configurable_2023} and creates executable code as a quantum circuit or quantum assembly code. This work aims to minimize the time for quantum state entanglement detection, thereby optimizing the code generated during quantum state initialization by a compiler. The proposed method builds upon Ref.~\cite{PhysRevA.71.052330}, where each level of the abstract syntax tree represents a quantum multiplexer. Therefore, one way to increase the efficiency of this procedure involves the analysis of the syntax tree to obtain reduced or parallel multiplexers. 
\subsection{Contributions}
We make the following contributions.
\begin{itemize}
    \item \textbf{Multiplexer simplification}. Our first result is the simplification of quantum multiplexers with the elimination of unnecessary controlled operations. Each removed control eliminates half of the multiplexer gates. If all controls are removed, the multiplexer circuit depth is reduced to one. The multiplexer simplification has linearithmic complexity. 
    \item \textbf{State preparation}. The initialization of unentangled states with multiplexers produces multiplexers with repeated operators. In this case, we can reduce the multiplexer's size and produce circuits with a reduced depth, the depth will be $O(2^{n_e})$, where $n_e$ is the number of qubits of the larger entangled component of the state.  Through an empirical evaluation, we verified that the depth of the circuits produced by the proposed model is competitive with the BAA algorithm~\cite{10.1109/TCAD.2023.3297972}.
\end{itemize}

\textbf{Evaluation}. We evaluated the proposed method to initialize quantum states with 4 to 12 qubits. For states with twelve qubits, the time to compile unentangled states with the proposed method is reduced by at least an order of magnitude compared with previous state preparation algorithms. In relation to the circuit depth, the proposed method is competitive with the BAA approach~\cite{10.1109/TCAD.2023.3297972}, which requires a search in the space of qubits.

The rest of this paper has four sections. In Section~\ref{sec:qc}, we introduce basic concepts about quantum computing. Section~\ref{sec:spm} describes previous procedures for quantum state preparation. Section~\ref{sec:mux_simp} is the main section, where we present an optimization to prepare unentangled quantum states more efficiently. The proposed method optimizes quantum multiplexers. Section~\ref{sec:conclusion} presents the conclusion and possible future work.

\section{Quantum Circuits} \label{sec:qc}
\subsection{Qubits}

Quantum bits, also known as qubits, are the basic units of quantum computing and are described in a bidimensional complex vector space. There is an infinite continuous set of possible values for a qubit (as opposed to the discrete set ${\{0, 1\}}$ that encompasses all the possible values for a classical bit). Each vector in this space is described by a linear combination of the basis vectors. Usually, we use $\{\ket{0},  \ket{1}\}$ as the basis, where:

$$\ket{0} = \left[ \begin{array}{c}1\\ 0 \end{array} \right] \mbox{ and } \ket{1} = \left[ \begin{array}{c}0\\1\end{array} \right].$$

A quantum state is the mathematical description of the information stored in a quantum system.  In some cases, a state $\ket{\psi}$ is a non-trivial linear combination of two orthonormal states $\ket{s_1}$ and $\ket{s_2}$. In this case, we have a superposition of $\ket{s_1} $ and $ \ket{s_2}$. For example, states $\ket{+} = \frac{1}{\sqrt{2}}\ket{0} + \frac{1}{\sqrt{2}}\ket{1}$ and $\ket{-} = \frac{1}{\sqrt{2}}\ket{0} - \frac{1}{\sqrt{2}}\ket{1}$, which are in a superposition concerning the standard computational basis.

The postulates of quantum mechanics tell us that measuring a state $\ket{\psi} = a\ket{s_1}+b\ket{s_2}$ has a probability $|a|^2$ of results in $\ket{s_1}$ and $|b|^2$ of results in $\ket{s_2}$. As expected, $|a|^2 +  |b|^2 = 1$.  Therefore, while a qubit can exist in a superposition, we can only extract one classical bit of information from it.

Describing a state with $n$ qubits in a classical device requires $2^n$-dimensional vectors. For example, a possible state with two qubits can be described by

$$a\ket{00}+b\ket{01} + c\ket{10} + d\ket{11}  = \left[ \begin{array}{c}a\\ b \\ c \\ d \end{array} \right]. $$

A quantum state $\ket{\psi}$ with more than one quantum register is entangled if it cannot be described as the tensor product of its components. 
For instance, the state $\frac{1}{\sqrt{2}}(\ket{00}+\ket{11})$ is entangled and has practical implications in the quantum teleportation protocol.
A quantum state $\ket{\psi} \in \mathcal{H}$ is unentangled for a given tensor decomposition $\mathcal{H}= \mathcal{H}_1 \otimes \cdots \otimes \mathcal{H}_n$ if $\ket{\psi}= \ket{\psi_1}\otimes \cdots \otimes \ket{\psi_n}$ with $\ket{\psi_i} \in \mathcal{H}_i$. In the next sections, we name unentangled state a state that is unentangled in relation to some tensor decomposition.

\subsection{State Transformations}
A quantum state transformation is a mapping from the quantum state space to itself. These transformations are described by a unitary operator. An operator $U:\mathcal{H}\mapsto \mathcal{H}$ in a complex vector space $\mathcal{H}$ is called unitary if its inverse is equivalent to its adjoint, i.e., $U^{\dag}U = UU^{\dag} = I$. All of these transformations can be seen as an operation in the complex vector space associated with the state space of a qubit.

An important example of quantum gates is the set of Pauli matrices:
$$X = \begin{pmatrix} 0 & 1\\ 1 & 0 \end{pmatrix},
Y = \begin{pmatrix} 0 & -i\\ i & 0 \end{pmatrix} \mbox{, and }
Z = \begin{pmatrix} 1 & 0\\ 0 & -1 \end{pmatrix}.$$
These gates represent $\pi$ rotations around each of the Bloch sphere's axes~\cite{Nielsen_Chuang_2010}.

Rotation operators are a common set of quantum gates that allow parameterized rotations around each one of the axes. These operators are
$$R_X(\theta) = \begin{pmatrix} \cos{\frac{\theta}{2}}  & -i \sin{\frac{\theta}{2}} \\ -i \sin{\frac{\theta}{2}}  & \cos{\frac{\theta}{2}}  \end{pmatrix},
R_Y(\theta) = \begin{pmatrix} \cos{\frac{\theta}{2}} & -\sin{\frac{\theta}{2}}\\ \sin{\frac{\theta}{2}} & \cos{\frac{\theta}{2}} \end{pmatrix}, \mbox{ and }
R_Z(\theta) = \begin{pmatrix} e^{-i \frac{\theta}{2}} & 0\\ 0 & e^{i \frac{\theta}{2}} \end{pmatrix}.$$

The Hadamard transformation that maps $\ket{0} \mapsto \ket{+}$ and $\ket{1} \mapsto \ket{-}$ creating a superposition of $\ket{0}$ and $\ket{1}$ is represented by

$$H = \dfrac{1}{\sqrt{2}} \begin{pmatrix} 1 & 1\\ 1 & -1 \end{pmatrix}.$$

Operations on multiple qubits can be constructed by taking the tensor product of single-qubit operators. For example, the operator $H \otimes H$ applies the Hadamard gate $H$ independently to both the first and second qubits. However, not all multi-qubit operations can be expressed as simple tensor products. A key example is the controlled-NOT (CNOT) gate, which acts on two qubits by flipping the second qubit if the first qubit is in the state $\ket{1}$. Its matrix representation is as follows: 

$$CNOT = \begin{pmatrix} 1 & 0 & 0 & 0\\ 0 & 1 & 0 & 0 \\ 0 & 0 & 0 & 1\\ 0 & 0 & 1 & 0\end{pmatrix}.$$

\subsection{Quantum Circuits}

It is possible to describe qubits and their transformations using a notation similar to that of classical digital systems. In these circuit diagrams, a horizontal wire represents a quantum register, while operators appear as rectangular boxes. The information flows from left to right. For example, a simple circuit with one qubit and an arbitrary operator $U$ is given by 
\begin{center}
\includegraphics{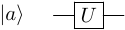}
\end{center}
which corresponds to $U\ket{a}$.

The CNOT gate acting on a two-qubit system is depicted as
\begin{center}
\includegraphics{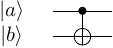}
\end{center}
expressing the operation $CNOT\ket{ab}=\ket{a}X^{a}\ket{b}$.

More generally, any operator controlled by $n$-qubits and acting on a single target qubit can be illustrated as follows.
\begin{center}
\includegraphics{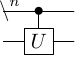}
\end{center}

\subsection{Quantum Multiplexers}

\begin{figure}[!htbp]
\includegraphics{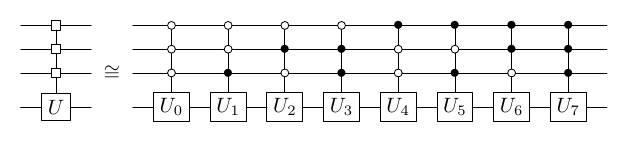}
\caption{Quantum multiplexer with three control qubits. The multiplexer can be represented by a list with the multicontrolled gates $U_j$. The $k$th control of $U_j$ is open (closed) when the $k$th bit in the binary representation of $j$ is equal to zero (one)\label{fig:mux3}.}
\Description{}
\end{figure}

A quantum multiplexer implements a conditional structure in a quantum circuit~\cite{1629135}. A multiplexer $U^{k}_{t}$ applies a unitary operator between a set of operators $\{U_0, U_1, ..., U_{N-1}\}$ on a target qubit $t$ according to the values of a set of $k$ qubits that act as controls, where $N = 2^k$. 
If the target $t$ is the most significative qubit, the matrix representation of a multiplexer can be viewed as a block diagonal matrix:
$$
    U^{k}_{t} = \begin{pmatrix} U_0 & & & \\ & U_1 & & \\ & & ... & \\ & & & U_{N-1}\end{pmatrix}.
$$
In quantum circuits, we represent a multiplexer control with a square symbol and it is equivalent to a sequence of multicontrolled gates, where a closed control (black circle) represents that the operator is applied when the control bit is equal to $\ket{1}$, and 
open control (white circle) represents that the operator is applied when the control bit is equal to $\ket{0}$. The multiplexer with three controls is exemplified in Fig.~\ref{fig:mux3}.

\begin{figure}[!htbp]
\includegraphics{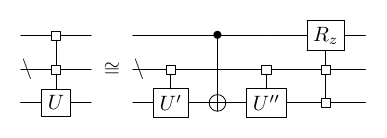}
\caption{Decomposition of a quantum multiplexer\label{fig:muxdec}}
\Description{}
\end{figure}

An efficient way to implement multiplexers is described in Fig.~\ref{fig:muxdec}~\cite{PhysRevA.71.052330}.
A multiplexer with $N$ operators produces a circuit with $O(N)$ depth and $O(log(N))$ qubits. Multiplexers are a basic building block of quantum algorithms and are used in the synthesis of arbitrary quantum gates~\cite{1629135, iten2016quantum}, quantum state preparation~\cite{iten2016quantum,1629135, PhysRevA.71.052330} and data initialization in quantum machine learning~\cite{blank2022compact}.

\section{Quantum State Preparation with Multiplexers} \label{sec:spm}

The problem of quantum state preparation (QSP) involves encoding classical data into a quantum state, which is a fundamental step for various quantum algorithms. An approach to achieve this is finding a multiplexer that transforms the initial state $\ket{a}_n$ into $\ket{a'}_{n-1}\ket{0}$~\cite{PhysRevA.71.052330} and then recursively applies a sequence of multiplexers to convert $\ket{a}_n$ into $\ket{0}_n$. Once this circuit is determined, its inverse can be used to convert the initial state $\ket{0}_n$ into the initial state.

Given an $n$ qubit state $\ket{\psi}=\sum_{k=0}^{2^n-1} a_k e^{iw_k} \ket{k}$, 
to determine the multiplexer gates $U_j$, we rewrite 
$$\ket{\psi}=\sum_{j=0}^{2^{n-1}-1}\sqrt{a_{2j}^2+a_{2j+1}^2 } \ket{j} \left(\frac{a_{2j} e^{iw_{2j}} \ket{0}+a_{2j+1} e^{iw_{2j+1} } \ket{1}}{\sqrt{a_{2j}^2+a_{2j+1}^2}}\right)$$ 
and for each $j$ we compute $\gamma_j=a_{2j+1}/\sqrt{a_{2j}^2+a_{2j+1}^2 }$ and apply  $U_j=R_y (-2\arcsin(\gamma_j)) R_z (w_{2j}-w_{2j+1} )$ in the rightmost qubit $r$, if the control qubits are equal to $j$. 
The action of the multiplexer $U_r^{n-1}$ with operators $\left\{U_j \right\}_{j=0}^{2^{n-1}-1}$ separates the rightmost qubit of the rest of the state.

$$U_r^{n-1} \ket{\psi}=\sum_{j=0}^{2^{n-1}-1}\sqrt{a_{2j}^2+a_{2j+1}^2 } e^{i(w_{2j}+w_{2j+1} )/2} \ket{j}\otimes\ket{0}$$

\begin{figure}[h]
\centering
\subfloat[]{\includegraphics[width=0.9\textwidth]{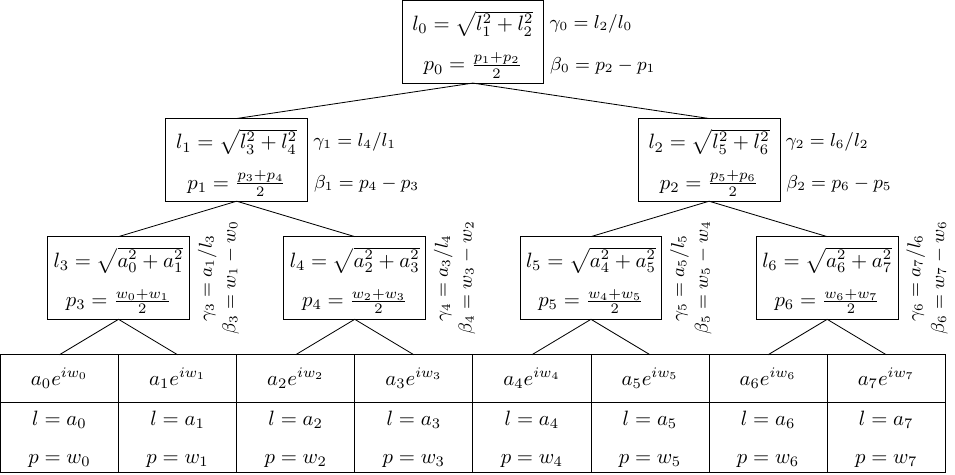} \label{fig:a}} \\
\subfloat[]{\includegraphics{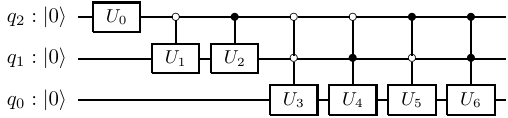}} \label{fig:b}
\caption{(a) Abstract tree representing a state preparation with quantum multiplexers. (b) Quantum state preparation circuit produced with the abstract tree, where $U_k = R_z(\beta_k)R_y(2\arcsin(\gamma_k))$. \label{fig:tree_hi}}
\Description{}
\end{figure}

One can use a binary abstract tree to compute the sequence of multiplexers used for state preparation. 
Level $k$ of the tree contains $2^k$ nodes representing a size $2^k$ multiplexer, as illustrated in Fig.~\ref{fig:tree_hi}. 
Leaf nodes have attributes value, phase and length and inner nodes have attributes phase $p$, length $l$, $\gamma$ (angle for the $y$-rotation), $\beta$ (angle for the $z$-rotation), and childs $left$ and $right$. The $value$ of the leafs corresponds to complex numbers of the state to be initialized. In the leaves, the $length$ attribute is equal to $|value|$ and $phase$ is equal to the phase of the number $value$ (in the interval $[-\pi, \pi]$, to avoid a difference of phases of equivalent operators). For each node that is not a leaf, the length is equal to $\sqrt{left.length^2+right.length^2}$, the phase attribute is equal to $(left.phase + right.phase)/2$, $\gamma = right.length / node.length$ and $\beta=right.phase - left.phase$.
The initialization of a quantum register with $n$ qubits requires $n$ multiplexers with $n-1, n-2, \cdots, 0$ controls. A $k$-controlled multiplexer produces a circuit with depth proportional to $2^k$, then the state preparation circuit has depth proportional to $\sum_{k=0}^{n-1}2^k=2^n$.

\section{Multiplexers Simplification} \label{sec:mux_simp}

\subsection{Optimization Method}

\begin{figure}[!htbp]%
\includegraphics{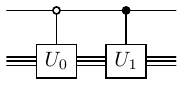}%
\caption{One controlled multiplexer with several targets.}%
\label{fig:onectrl_mux}%
\Description{}
\end{figure}

There is a relationship between state separation and the presence of repetition patterns in multiplexers. First consider a multiplexer with one control qubit as in Fig.~\ref{fig:onectrl_mux}. With input $(a\ket{0} + b\ket{1})\ket{0}$, this operator creates the state $a\ket{0}U_0\ket{0} + b\ket{1}U_1\ket{0}$. If $U_1 = e^{i\alpha}U_0$ the circuit does not entangle the two qubits because $a\ket{0}U_0\ket{0} + b\ket{1}U_1\ket{0}= (a\ket{0} + be^{i\alpha}\ket{1})U_0\ket{0}$.

\begin{figure}[!htbp]
    \centering
		\includegraphics[width=\textwidth]{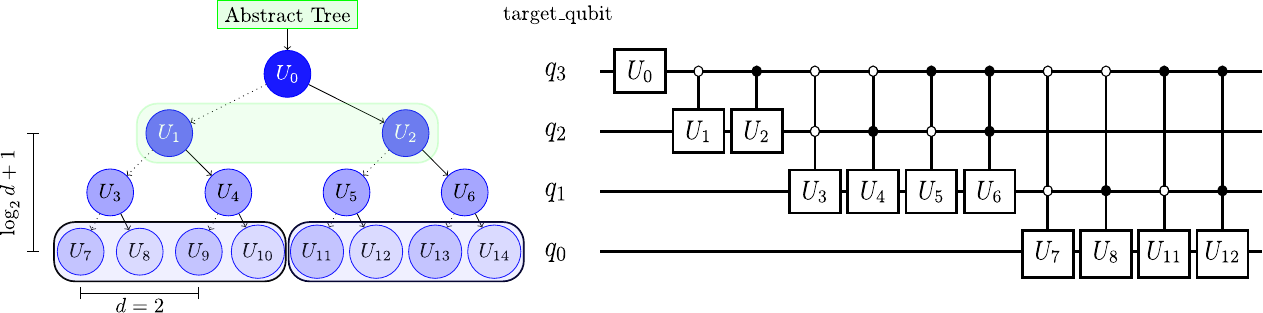}

    \caption{  (left) Tree of operators to initialize a quantum state with multiplexers. Each level of the tree defines a multiplexer that is stored in an array. Dotted (solid) arrows correspond to open (closed) controls of the gate in the multiplexer.  (right) Circuit corresponding to the state preparation circuit if $[U_7, U_8]=[U_9, U_{10}]$ and $[U_{11}, U_{12}]=[U_{13}, U_{14}]$.}
    \label{fig:angle_tree}
    \Description{}
\end{figure}

Consider the tree of operators in Fig.~\ref{fig:angle_tree}, where each layer represents a multiplexer and $U_k=R_z(\beta_k)R_y(2arcsin(\gamma_k))$ ($\beta_k$ and $\gamma_i$ from Fig.~\ref{fig:tree_hi}). Dashed edges correspond to open controls and solid edges correspond to close controls.
If the leaves of the subtree with roots $U_3$ and $U_4$ are equal and the leaves of the subtree with roots $U_5$ and $U_6$ are also equal then we can remove the controls in qubit $q_2$ of the last multiplexer. In this case, $[U_7, U_8]=[U_9, U_{10}]$ and $[U_{11}, U_{12}]=[U_{13}, U_{14}]$. The boxes with size $2d = 4$ separate the leaves of the subtrees with roots $U_1$ and $U_2$, inside each box if $U_j$ and $U_{j+d}$ are equal, then the controls in $q_2$ of the last multiplexer can be removed, as shown in the circuit in Fig.~\ref{fig:angle_tree}.

In a multiplexer with $n$ controls, if we can split the multiplexer gates into $2^k$ sequential partitions with the same size $2d=2^n/2^k$ (corresponding to leaves of subtrees of height $\log_2(d)+1$) and in each partition the $j$th gate is equal to the $(j+d)$th gate, then 
the multiplexer controls in the qubit $(k + target\_bit)$ can be removed.
By leveraging these repetition patterns to identify state separability, multiplexer simplification can produce quantum circuits with reduced depth and fewer CNOT gates.

Algorithm~\ref{alg:1} shows the \simplify function that implements the proposed multiplexer simplification. The operators of each multiplexer (level of abstract tree) are stored in an array. Line~\ref{line:rep_search} identifies repetitions in the multiplexer and returns the unnecessary controls and the indexes of the repeated operators that must be removed of the multiplexer. If the set deleted\_operators is not empty, Line~\ref{line:remove_ops} creates the simplified\_mux removing the unnecessary operators of mux. Algorithm 1 returns the unnecessary controls and the simplified multiplexer. 

\begin{algorithm}[!htbp]
\SetKwProg{Def}{def}{:}{}
\caption{\textbf{Simplify Function}}
\label{alg:1}
\SetAlgoNlRelativeSize{0}
\SetKwFunction{RepetitionSearch}{repetition\_search}
\SetKwFunction{lg}{log2}
\SetKwFunction{len}{len}
\SetKwFunction{set}{set}
\SetKwFunction{alist}{list}

\Def{\simplify(mux, target\_qubit)}{
    deleted\_operators = \set() \\
    removed\_controls = \alist() \\
    \BlankLine
    
    \If{\len(mux) > 1}{
        removed\_controls, deleted\_operators = \RepetitionSearch(mux, target\_qubit) \label{line:rep_search}
    }
    \BlankLine
    
    \If{deleted\_operators is not empty}{
        simplified\_mux = mux without indexes in deleted\_operators \label{line:remove_ops}\\
        \Return {removed\_controls, simplified\_mux}
    }
    \BlankLine
    
    simplified\_mux = mux \\
    \Return{removed\_controls, simplified\_mux}\;
}
\end{algorithm}

\begin{algorithm}[!htbp]
\SetKwProg{Def}{def}{:}{}
\caption{\textbf{RepetitionSearch Function}}
\label{alg:2}
\SetAlgoNlRelativeSize{0}

\Def{\RepetitionSearch(mux, target\_qubit)}{
	deleted\_operators = \set() \\
	deleted\_controls = \alist() \\

	\For{j = 0 $\KwTo$ \lg(\len(mux)) \label{line:for_power2}}{
		d = 2$^j$ \\
		delete\_set = \set() \\
		\If{mux[0] == mux[d]}{delete\_set = \RepetitionVerify(d, mux) \label{line:call_repetition_verify}}
		
		\If{delete\_set is not empty}{
			remove\_control = target\_qubit + \lg(d) + 1 \\
			Add removed\_control into deleted\_controls \\
			Add deleted\_set into deleted\_operators
		}
	}

	\Return{deleted\_controls, deleted\_operators}
}
\end{algorithm}

Algorithm~\ref{alg:2} describes \RepetitionSearch function that receives a multiplexer and the index of target qubit of the multiplexer (for instance, see Fig.~\ref{fig:angle_tree}, where the target\_qubit of each multiplexer is at the right side of the multiplexer gates). The positions of the multiplexer whose indexes $d$ are a power of two are traveled by the loop in line~\ref{line:for_power2} to find an operator identical to the first one (mux[0] == mux[d]). If a repeated operator is found, Line~\ref{line:call_repetition_verify} call the function \RepetitionVerify that returns the set deleted\_set with indexes of the unnecessary operators. If delete\_set is not empty, lines 10-12 include target\_qubit + \lg(d) + 1 in the list removed\_controls and the values in delete\_set into deleted\_controls.

The \RepetitionVerify function verifies if the multiplexer mux can be split into $len(mux) / (2 * d)$ sequential partitions, where
the operators in the first half of each partition are equal to the second half. If the two halves of all partitions are equal, the list of
indexes of the second half of the partitions are returned.

The total cost of the multiplexer simplification is linearithmic in terms of the size of the multiplexer. In the worst case, Algorithm \ref{alg:1} calls \RepetitionSearch one time and the removal of marked operators has a linear cost. So, the total cost for the simplification of one multiplexer is given by $O(N + C_{repetition\_search})$, where $C_{repetition\_search}$ is the cost of function \RepetitionSearch in Algorithm~\ref{alg:2}. The call to the function in Algorithm \ref{alg:2} is the main contribution to the total time complexity, as explained below.

In Algorithm \ref{alg:2}, we traveled the positions of the array whose index is a power of two. So, the operations in the interior of the loop in Line~\ref{line:for_power2} are executed $O(\log_2{N})$ times. The function \RepetitionVerify traverses all the operators and has linear complexity. The set deleted\_set has at most $N/2$ indexes and lines 10 to 11 have also a linear complexity. The complexity of \RepetitionSearch is $C_{repetition\_search} =O(N \log_2{N}))$ and we conclude that the complexity of the overall simplification is linearithmic.

\subsection{Experiments and Discussion}

We compare the proposed multiplexer simplification method combined with the state preparation method from Ref.~\cite{PhysRevA.71.052330}, denoted as UCGE (an acronym for the union of Uniformly Controlled One-Qubit Gates with Entanglement), with other state preparation approaches~\cite{PhysRevA.71.052330, 10.1109/TCAD.2023.3297972, PhysRevA.83.032302}. 
To evaluate the efficiency of our simplification method for unentangled states, we conducted two experiments. 

\begin{figure}[t]
    \centering
    \subfloat[]{
        \includegraphics[width=0.49\linewidth]{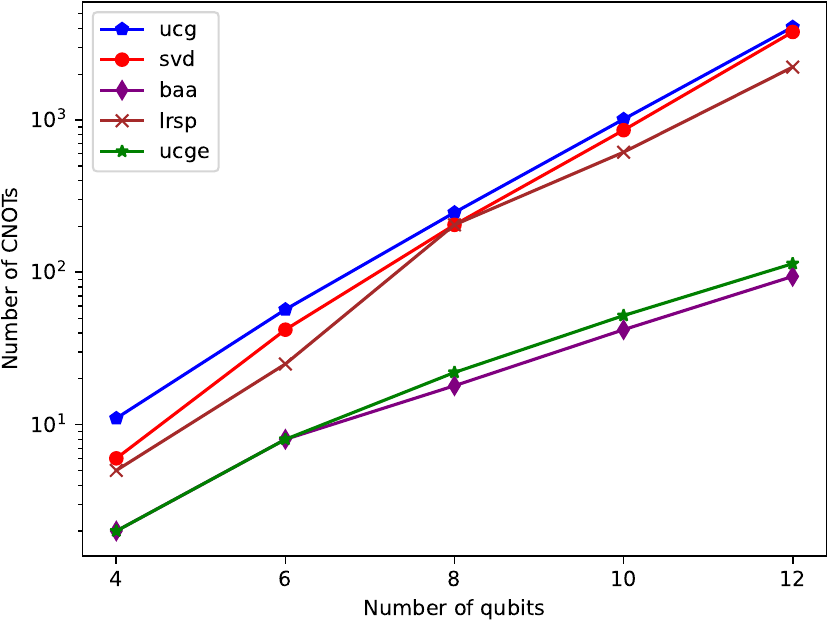} \label{fig:cnots_bipartition}
        }
    \subfloat[]{
        \includegraphics[width=0.49\linewidth]{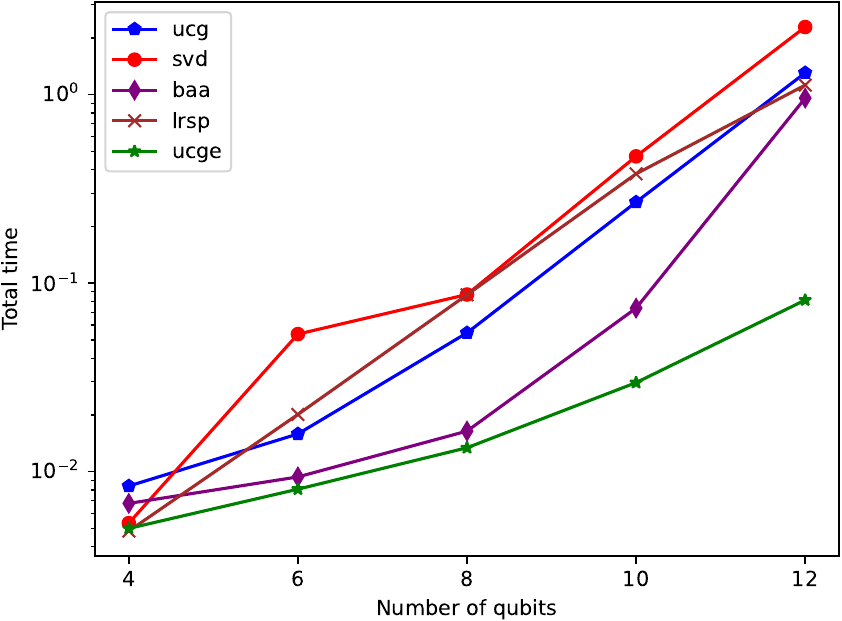} \label{fig:time_bipartition}
    }
    \caption{The number of CNOT gates in the circuit, along with the execution and transpilation times (in seconds), are compared for preparing a bipartite quantum state with varying numbers of qubits. The algorithms used for the comparison are: `ucg'~\cite{PhysRevA.71.052330}, `lrsp'~\cite{10.1109/TCAD.2023.3297972},
    `baa'~\cite{10.1109/TCAD.2023.3297972},
    `svd'~\cite{PhysRevA.83.032302}, and the proposed method `ucge'. The `ucg' algorithm is implemented in the Qiskit library~\cite{Qiskit}, while the others are implemented in the Qclib library~\cite{qclib}.
    }  \label{fig:graph1}
    \Description{}
\end{figure}

In the first experiment, we prepare a set of random complex-valued unentangled states varying the number of qubits $n$. Each random state $\ket{\psi}$ consists of two unentangled components $\ket{\psi} = \ket{\psi_0} \otimes \ket{\psi_1}$, where $\ket{\psi_0}$ and $\ket{\psi_1}$ have $2^{n/2}$ amplitudes from a uniform random distribution. We use the state $\ket{\psi}$ as the input of the state preparation methods. The number of CNOT gates in Fig.~\ref{fig:cnots_bipartition} is determined by the decomposition of the state preparation circuit into CNOT and single-qubit gates using the transpile function from qiskit version 1.4.0 with optimization level equal to 2. The time in Fig.~\ref{fig:time_bipartition} includes the time to create and transpile the circuit. Fig.~\ref{fig:graph1} illustrates that the total time for BAA increases at a rate of $N/\log_2(N)$ compared to UCGE in the bipartite state initialization. 

\begin{figure}[t]
    \centering
    \subfloat[]{
        \includegraphics[width=0.49\linewidth]{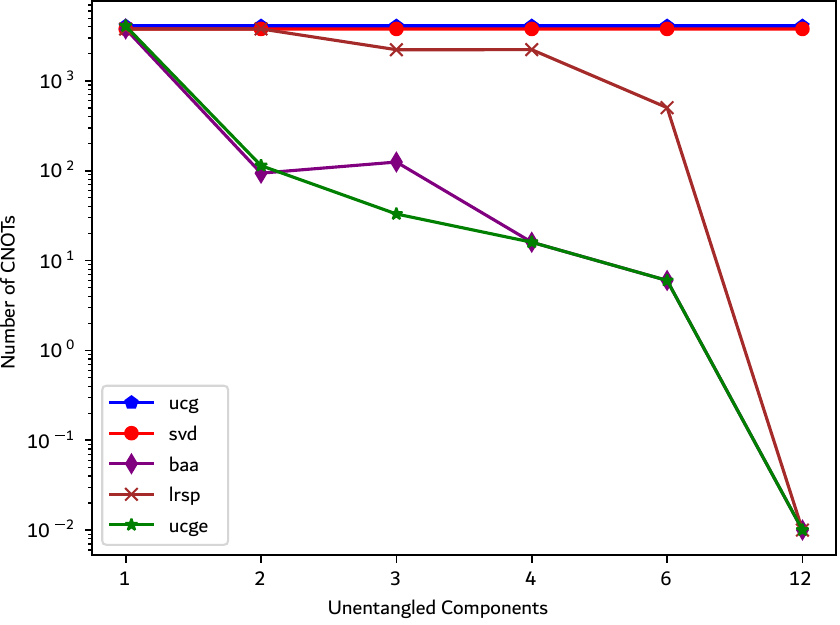} \label{fig:cnots_multipartite}
        }
    \subfloat[]{
        \includegraphics[width=0.49\linewidth]{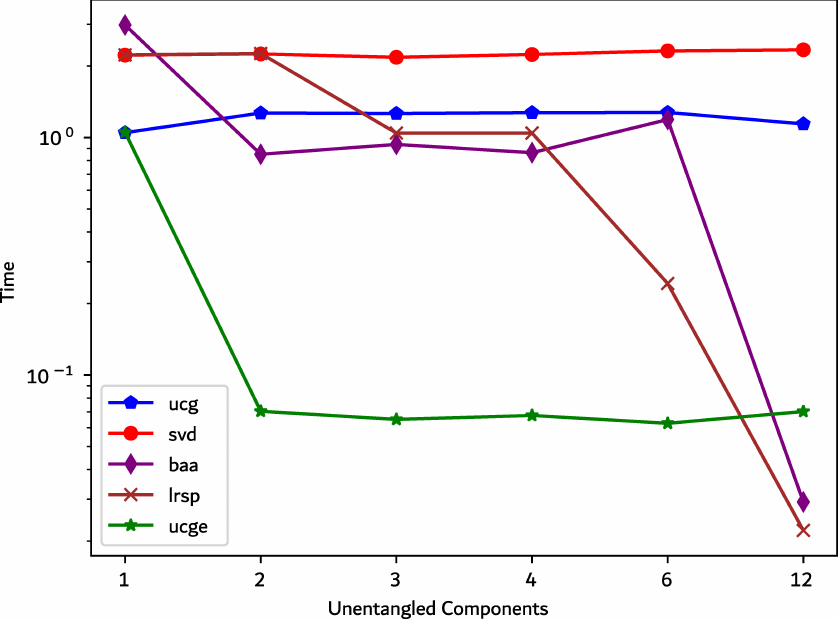} \label{fig:time_multipartite}
    }
    \caption{The number of CNOT gates in the circuit, along with the execution and transpilation times (in seconds), are compared to prepare an n-partite 12-qubit quantum state with equally sized partition blocks. The algorithms used for the comparison are: `ucg'~\cite{PhysRevA.71.052330}, `lrsp'~\cite{10.1109/TCAD.2023.3297972},
    `baa'~\cite{10.1109/TCAD.2023.3297972},
    `svd'~\cite{PhysRevA.83.032302}, and the proposed method `ucge'. The `ucg' algorithm is implemented in the Qiskit library~\cite{Qiskit}, while the others are implemented in the Qclib library~\cite{qclib}.
    }  \label{fig:graph2}
    \Description{}
\end{figure}

The second experiment fixes the number of qubits at twelve and varies the number of components $k$ of the unentangled state $|\psi\rangle = |\psi_1\rangle \otimes \cdots \otimes |\psi_k\rangle$, where each subspace has $12 / k$ qubits. The same QSP techniques are used for this experiment. The number of CNOT gates in Fig.~\ref{fig:cnots_multipartite} is determined by the decomposition of the state preparation circuit into CNOT and single-qubit gates using the transpile function from qiskit version 1.4.0 with optimization level equal to 2. The time in Fig.~\ref{fig:time_multipartite} includes the time to create and transpile the circuit. 
There is a clear advantage in the time of the proposed approach. However, in the completely separable case the BAA algorithm does not need to perform a search in the space of qubits and in this case the BAA has a time advantage.

The reduction in the number of controls on multiplexers explains why the QSP circuit has fewer CNOT gates compared to previous state preparation approaches that do not detect unentangled states~\cite{PhysRevA.71.052330, PhysRevA.83.032302, 1629135}. Without separability detection, the number of CNOT gates in the resulting circuit increases exponentially with the total number of qubits. Removing some controls significantly decreases the cost, as shown in Fig.~\ref{fig:graph1} and Fig.~\ref{fig:graph2}. In terms of exact state preparation, our results are highly competitive with the BAA method~\cite{10.1109/TCAD.2023.3297972}, which also employs separability detection for this purpose.

The main aspect that sets our approach apart from others is the efficient detection of unentangled states at a reduced computational cost. The proposed method demonstrates an advantage in the total time required to generate a state preparation circuit, including circuit generation and transpilation time. Transpilation into CNOT and unitary one-qubit operators is essential because quantum computers operate with a limited set of quantum gates. Our method's shorter QSP time compared to strategies that do not detect unentanglement is due to replacing a multiplexer with smaller multiplexers, which reduces the size of the transpiled circuit. As noted previously, the time for simplification search and application has a worst-case scenario of linearithmic complexity relative to the multiplexer size. Consequently, the impact of the search on the total QSP time is smaller than that of methods employing brute-force search~\cite{10.1109/TCAD.2023.3297972}. 

\section{Conclusion}\label{sec:conclusion}

The quantum multiplexer simplification for state preparation reduces the number of CNOT gates required for QSP, reducing the time complexity for state separation detection. This is accomplished by identifying repetitions in the abstract tree that describe the set of multiplexers implementing a quantum state. To demonstrate the utility of this approach, we compare our results with other techniques for encoding data into the amplitudes of a quantum state. The time advantage of the proposed method, which explores separability, is evident when compared with other state preparation algorithms. 

Future work could involve applying this optimization method to initialize approximated quantum states~\cite{rudolph2023decomposition, gundlapalli2022deterministic}. Optimize other state preparation techniques with the optimization of the abstract tree (or decision diagrams~\cite{wille2022decision, mozafari2022efficient}) and using the multiplexer optimization in various quantum encoding scenarios where multiplexers are needed~\cite{iten2016quantum, 1629135}, or separable states are common~\cite{gundlapalli2022deterministic}.

\section*{Data availability}
The data and software generated during this study are available at \url{https://github.com/qclib/qclib-papers} 
and \url{https://github.com/qclib/qclib}~\cite{qclib}. The simplification proposed in this work is applied by default in the UCGate of qiskit version 2.0.0rc1.

\section*{Conflict of interests}

All authors declare no conflicts of interest.

\section*{Acknowledgments}
 This work is supported by research grants from CNPq, CAPES and FACEPE (Brazilian research agencies).
 
\bibliographystyle{ACM-Reference-Format}

\begin{thebibliography}{32}

\ifx \showCODEN    \undefined \def \showCODEN     #1{\unskip}     \fi
\ifx \showDOI      \undefined \def \showDOI       #1{#1}\fi
\ifx \showISBNx    \undefined \def \showISBNx     #1{\unskip}     \fi
\ifx \showISBNxiii \undefined \def \showISBNxiii  #1{\unskip}     \fi
\ifx \showISSN     \undefined \def \showISSN      #1{\unskip}     \fi
\ifx \showLCCN     \undefined \def \showLCCN      #1{\unskip}     \fi
\ifx \shownote     \undefined \def \shownote      #1{#1}          \fi
\ifx \showarticletitle \undefined \def \showarticletitle #1{#1}   \fi
\ifx \showURL      \undefined \def \showURL       {\relax}        \fi
% The following commands are used for tagged output and should be
% invisible to TeX
\providecommand\bibfield[2]{#2}
\providecommand\bibinfo[2]{#2}
\providecommand\natexlab[1]{#1}
\providecommand\showeprint[2][]{arXiv:#2}

\bibitem[Araujo et~al\mbox{.}(2023a)]%
        {qclib}
\bibfield{author}{\bibinfo{person}{Israel~F. Araujo}, \bibinfo{person}{Ismael
  C.~S. Araújo}, \bibinfo{person}{Leon~D. da Silva}, \bibinfo{person}{Carsten
  Blank}, {and} \bibinfo{person}{Adenilton~J. da Silva}.}
  \bibinfo{year}{2023}\natexlab{a}.
\newblock \bibinfo{booktitle}{\emph{{Quantum computing library}}}.
\newblock
\urldef\tempurl%
\url{https://github.com/qclib/qclib}
\showURL{%
\tempurl}


\bibitem[Araujo et~al\mbox{.}(2023b)]%
        {10.1109/TCAD.2023.3297972}
\bibfield{author}{\bibinfo{person}{Israel~F. Araujo}, \bibinfo{person}{Carsten
  Blank}, \bibinfo{person}{Ismael C.~S. Ara\'{u}jo}, {and}
  \bibinfo{person}{Adenilton~J. da Silva}.} \bibinfo{year}{2023}\natexlab{b}.
\newblock \showarticletitle{Low-Rank Quantum State Preparation}.
\newblock \bibinfo{journal}{\emph{Trans. Comp.-Aided Des. Integ. Cir. Sys.}}
  \bibinfo{volume}{43}, \bibinfo{number}{1} (\bibinfo{date}{July}
  \bibinfo{year}{2023}), \bibinfo{pages}{161–170}.
\newblock
\showISSN{0278-0070}
\urldef\tempurl%
\url{https://doi.org/10.1109/TCAD.2023.3297972}
\showDOI{\tempurl}


\bibitem[Araujo et~al\mbox{.}(2023c)]%
        {araujo_configurable_2023}
\bibfield{author}{\bibinfo{person}{Israel~F. Araujo},
  \bibinfo{person}{Daniel~K. Park}, \bibinfo{person}{Teresa~B. Ludermir},
  \bibinfo{person}{Wilson~R. Oliveira}, \bibinfo{person}{Francesco
  Petruccione}, {and} \bibinfo{person}{Adenilton~J. da Silva}.}
  \bibinfo{year}{2023}\natexlab{c}.
\newblock \showarticletitle{Configurable sublinear circuits for quantum state
  preparation}.
\newblock \bibinfo{journal}{\emph{Quantum Information Processing}}
  \bibinfo{volume}{22}, \bibinfo{number}{2} (\bibinfo{year}{2023}),
  \bibinfo{pages}{123}.
\newblock
\showISSN{1573-1332}
\urldef\tempurl%
\url{https://doi.org/10.1007/s11128-023-03869-7}
\showDOI{\tempurl}


\bibitem[Araujo et~al\mbox{.}(2021)]%
        {araujo_divide_2021}
\bibfield{author}{\bibinfo{person}{Israel~F Araujo}, \bibinfo{person}{Daniel~K
  Park}, \bibinfo{person}{Francesco Petruccione}, {and}
  \bibinfo{person}{Adenilton~J da Silva}.} \bibinfo{year}{2021}\natexlab{}.
\newblock \showarticletitle{A divide-and-conquer algorithm for quantum state
  preparation}.
\newblock \bibinfo{journal}{\emph{Scientific reports}} \bibinfo{volume}{11},
  \bibinfo{number}{1} (\bibinfo{year}{2021}), \bibinfo{pages}{6329}.
\newblock


\bibitem[Benioff(1982)]%
        {benioff1982quantum}
\bibfield{author}{\bibinfo{person}{Paul Benioff}.}
  \bibinfo{year}{1982}\natexlab{}.
\newblock \showarticletitle{Quantum mechanical Hamiltonian models of Turing
  machines}.
\newblock \bibinfo{journal}{\emph{Journal of Statistical Physics}}
  \bibinfo{volume}{29} (\bibinfo{year}{1982}), \bibinfo{pages}{515--546}.
\newblock


\bibitem[Bergholm et~al\mbox{.}(2005)]%
        {PhysRevA.71.052330}
\bibfield{author}{\bibinfo{person}{Ville Bergholm}, \bibinfo{person}{Juha~J.
  Vartiainen}, \bibinfo{person}{Mikko M\"ott\"onen}, {and}
  \bibinfo{person}{Martti~M. Salomaa}.} \bibinfo{year}{2005}\natexlab{}.
\newblock \showarticletitle{Quantum circuits with uniformly controlled
  one-qubit gates}.
\newblock \bibinfo{journal}{\emph{Phys. Rev. A}}  \bibinfo{volume}{71}
  (\bibinfo{date}{May} \bibinfo{year}{2005}), \bibinfo{pages}{052330}.
\newblock
Issue 5.
\urldef\tempurl%
\url{https://doi.org/10.1103/PhysRevA.71.052330}
\showDOI{\tempurl}


\bibitem[Blank et~al\mbox{.}(2022)]%
        {blank2022compact}
\bibfield{author}{\bibinfo{person}{Carsten Blank}, \bibinfo{person}{Adenilton~J
  da Silva}, \bibinfo{person}{Lucas~P de Albuquerque},
  \bibinfo{person}{Francesco Petruccione}, {and} \bibinfo{person}{Daniel~K
  Park}.} \bibinfo{year}{2022}\natexlab{}.
\newblock \showarticletitle{Compact quantum kernel-based binary classifier}.
\newblock \bibinfo{journal}{\emph{Quantum Science and Technology}}
  \bibinfo{volume}{7}, \bibinfo{number}{4} (\bibinfo{year}{2022}),
  \bibinfo{pages}{045007}.
\newblock


\bibitem[Feynman(1982)]%
        {Feynman1982}
\bibfield{author}{\bibinfo{person}{Richard~P. Feynman}.}
  \bibinfo{year}{1982}\natexlab{}.
\newblock \showarticletitle{Simulating physics with computers}.
\newblock \bibinfo{journal}{\emph{International Journal of Theoretical
  Physics}} \bibinfo{volume}{21}, \bibinfo{number}{6} (\bibinfo{date}{June}
  \bibinfo{year}{1982}), \bibinfo{pages}{467--488}.
\newblock
\showISSN{1572-9575}
\urldef\tempurl%
\url{https://doi.org/10.1007/BF02650179}
\showDOI{\tempurl}


\bibitem[Gleinig and Hoefler(2021)]%
        {gleinig2021efficient}
\bibfield{author}{\bibinfo{person}{Niels Gleinig} {and}
  \bibinfo{person}{Torsten Hoefler}.} \bibinfo{year}{2021}\natexlab{}.
\newblock \showarticletitle{An Efficient Algorithm for Sparse Quantum State
  Preparation}. In \bibinfo{booktitle}{\emph{2021 58th ACM/IEEE Design
  Automation Conference (DAC)}}. \bibinfo{pages}{433--438}.
\newblock
\urldef\tempurl%
\url{https://doi.org/10.1109/DAC18074.2021.9586240}
\showDOI{\tempurl}


\bibitem[Grover(2000)]%
        {grover2000synthesis}
\bibfield{author}{\bibinfo{person}{Lov~K Grover}.}
  \bibinfo{year}{2000}\natexlab{}.
\newblock \showarticletitle{Synthesis of quantum superpositions by quantum
  computation}.
\newblock \bibinfo{journal}{\emph{Physical review letters}}
  \bibinfo{volume}{85}, \bibinfo{number}{6} (\bibinfo{year}{2000}),
  \bibinfo{pages}{1334}.
\newblock


\bibitem[Gui et~al\mbox{.}(2024)]%
        {gui2024spacetime}
\bibfield{author}{\bibinfo{person}{Kaiwen Gui}, \bibinfo{person}{Alexander~M
  Dalzell}, \bibinfo{person}{Alessandro Achille}, \bibinfo{person}{Martin
  Suchara}, {and} \bibinfo{person}{Frederic~T Chong}.}
  \bibinfo{year}{2024}\natexlab{}.
\newblock \showarticletitle{Spacetime-efficient low-depth quantum state
  preparation with applications}.
\newblock \bibinfo{journal}{\emph{Quantum}}  \bibinfo{volume}{8}
  (\bibinfo{year}{2024}), \bibinfo{pages}{1257}.
\newblock


\bibitem[Gundlapalli and Lee(2022)]%
        {gundlapalli2022deterministic}
\bibfield{author}{\bibinfo{person}{Prithvi Gundlapalli} {and}
  \bibinfo{person}{Junyi Lee}.} \bibinfo{year}{2022}\natexlab{}.
\newblock \showarticletitle{Deterministic and entanglement-efficient
  preparation of amplitude-encoded quantum registers}.
\newblock \bibinfo{journal}{\emph{Physical Review Applied}}
  \bibinfo{volume}{18}, \bibinfo{number}{2} (\bibinfo{year}{2022}),
  \bibinfo{pages}{024013}.
\newblock


\bibitem[Hughes et~al\mbox{.}(1996)]%
        {PhysRevLett.77.3240}
\bibfield{author}{\bibinfo{person}{Richard~J. Hughes}, \bibinfo{person}{Daniel
  F.~V. James}, \bibinfo{person}{Emanuel~H. Knill}, \bibinfo{person}{Raymond
  Laflamme}, {and} \bibinfo{person}{Albert~G. Petschek}.}
  \bibinfo{year}{1996}\natexlab{}.
\newblock \showarticletitle{Decoherence Bounds on Quantum Computation with
  Trapped Ions}.
\newblock \bibinfo{journal}{\emph{Phys. Rev. Lett.}}  \bibinfo{volume}{77}
  (\bibinfo{date}{Oct.} \bibinfo{year}{1996}), \bibinfo{pages}{3240--3243}.
\newblock
Issue 15.
\urldef\tempurl%
\url{https://doi.org/10.1103/PhysRevLett.77.3240}
\showDOI{\tempurl}


\bibitem[Iten et~al\mbox{.}(2016)]%
        {iten2016quantum}
\bibfield{author}{\bibinfo{person}{Raban Iten}, \bibinfo{person}{Roger
  Colbeck}, \bibinfo{person}{Ivan Kukuljan}, \bibinfo{person}{Jonathan Home},
  {and} \bibinfo{person}{Matthias Christandl}.}
  \bibinfo{year}{2016}\natexlab{}.
\newblock \showarticletitle{Quantum circuits for isometries}.
\newblock \bibinfo{journal}{\emph{Physical Review A}} \bibinfo{volume}{93},
  \bibinfo{number}{3} (\bibinfo{year}{2016}), \bibinfo{pages}{032318}.
\newblock


\bibitem[Javadi-Abhari et~al\mbox{.}(2024)]%
        {Qiskit}
\bibfield{author}{\bibinfo{person}{Ali Javadi-Abhari}, \bibinfo{person}{Matthew
  Treinish}, \bibinfo{person}{Kevin Krsulich}, \bibinfo{person}{Christopher~J.
  Wood}, \bibinfo{person}{Jake Lishman}, \bibinfo{person}{Julien Gacon},
  \bibinfo{person}{Simon Martiel}, \bibinfo{person}{Paul~D. Nation},
  \bibinfo{person}{Lev~S. Bishop}, \bibinfo{person}{Andrew~W. Cross},
  \bibinfo{person}{Blake~R. Johnson}, {and} \bibinfo{person}{Jay~M. Gambetta}.}
  \bibinfo{year}{2024}\natexlab{}.
\newblock \bibinfo{title}{Quantum computing with {Q}iskit}.
\newblock
\newblock
\urldef\tempurl%
\url{https://doi.org/10.48550/arXiv.2405.08810}
\showDOI{\tempurl}
\showeprint[arxiv]{2405.08810}~[quant-ph]


\bibitem[Long and Sun(2001)]%
        {long2001efficient}
\bibfield{author}{\bibinfo{person}{Gui-Lu Long} {and} \bibinfo{person}{Yang
  Sun}.} \bibinfo{year}{2001}\natexlab{}.
\newblock \showarticletitle{Efficient scheme for initializing a quantum
  register with an arbitrary superposed state}.
\newblock \bibinfo{journal}{\emph{Physical Review A}} \bibinfo{volume}{64},
  \bibinfo{number}{1} (\bibinfo{year}{2001}), \bibinfo{pages}{014303}.
\newblock


\bibitem[M{\"o}tt{\"o}nen et~al\mbox{.}(2005)]%
        {mottonen2005transformation}
\bibfield{author}{\bibinfo{person}{Mikko M{\"o}tt{\"o}nen}, \bibinfo{person}{JJ
  Vartiainen}, \bibinfo{person}{Ville Bergholm}, {and}
  \bibinfo{person}{Martti~M Salomaa}.} \bibinfo{year}{2005}\natexlab{}.
\newblock \showarticletitle{Transformation of quantum states using uniformly
  controlled rotations}.
\newblock \bibinfo{journal}{\emph{Quantum Information and Computation}}
  \bibinfo{volume}{5}, \bibinfo{number}{6} (\bibinfo{year}{2005}),
  \bibinfo{pages}{467--473}.
\newblock


\bibitem[Mozafari et~al\mbox{.}(2022)]%
        {mozafari2022efficient}
\bibfield{author}{\bibinfo{person}{Fereshte Mozafari},
  \bibinfo{person}{Giovanni De~Micheli}, {and} \bibinfo{person}{Yuxiang Yang}.}
  \bibinfo{year}{2022}\natexlab{}.
\newblock \showarticletitle{Efficient deterministic preparation of quantum
  states using decision diagrams}.
\newblock \bibinfo{journal}{\emph{Physical Review A}} \bibinfo{volume}{106},
  \bibinfo{number}{2} (\bibinfo{year}{2022}), \bibinfo{pages}{022617}.
\newblock


\bibitem[Nakaji et~al\mbox{.}(2022)]%
        {nakaji2022approximate}
\bibfield{author}{\bibinfo{person}{Kouhei Nakaji}, \bibinfo{person}{Shumpei
  Uno}, \bibinfo{person}{Yohichi Suzuki}, \bibinfo{person}{Rudy Raymond},
  \bibinfo{person}{Tamiya Onodera}, \bibinfo{person}{Tomoki Tanaka},
  \bibinfo{person}{Hiroyuki Tezuka}, \bibinfo{person}{Naoki Mitsuda}, {and}
  \bibinfo{person}{Naoki Yamamoto}.} \bibinfo{year}{2022}\natexlab{}.
\newblock \showarticletitle{Approximate amplitude encoding in shallow
  parameterized quantum circuits and its application to financial market
  indicators}.
\newblock \bibinfo{journal}{\emph{Physical Review Research}}
  \bibinfo{volume}{4}, \bibinfo{number}{2} (\bibinfo{year}{2022}),
  \bibinfo{pages}{023136}.
\newblock


\bibitem[Nielsen and Chuang(2010)]%
        {Nielsen_Chuang_2010}
\bibfield{author}{\bibinfo{person}{Michael~A. Nielsen} {and}
  \bibinfo{person}{Isaac~L. Chuang}.} \bibinfo{year}{2010}\natexlab{}.
\newblock \bibinfo{booktitle}{\emph{Quantum Computation and Quantum
  Information: 10th Anniversary Edition}}.
\newblock \bibinfo{publisher}{Cambridge University Press},
  \bibinfo{address}{Cambridge, UK}.
\newblock


\bibitem[Park et~al\mbox{.}(2019)]%
        {park2019circuit}
\bibfield{author}{\bibinfo{person}{Daniel~K Park}, \bibinfo{person}{Francesco
  Petruccione}, {and} \bibinfo{person}{June-Koo~Kevin Rhee}.}
  \bibinfo{year}{2019}\natexlab{}.
\newblock \showarticletitle{Circuit-based quantum random access memory for
  classical data}.
\newblock \bibinfo{journal}{\emph{Scientific reports}} \bibinfo{volume}{9},
  \bibinfo{number}{1} (\bibinfo{year}{2019}), \bibinfo{pages}{3949}.
\newblock


\bibitem[Plesch and Brukner(2011)]%
        {PhysRevA.83.032302}
\bibfield{author}{\bibinfo{person}{Martin Plesch} {and}
  \bibinfo{person}{\ifmmode \check{C}\else~\v{C}\fi{}aslav Brukner}.}
  \bibinfo{year}{2011}\natexlab{}.
\newblock \showarticletitle{Quantum-state preparation with universal gate
  decompositions}.
\newblock \bibinfo{journal}{\emph{Phys. Rev. A}}  \bibinfo{volume}{83}
  (\bibinfo{date}{March} \bibinfo{year}{2011}), \bibinfo{pages}{032302}.
\newblock
Issue 3.
\urldef\tempurl%
\url{https://doi.org/10.1103/PhysRevA.83.032302}
\showDOI{\tempurl}


\bibitem[Preskill(2018)]%
        {preskill2018quantum}
\bibfield{author}{\bibinfo{person}{John Preskill}.}
  \bibinfo{year}{2018}\natexlab{}.
\newblock \showarticletitle{Quantum computing in the NISQ era and beyond}.
\newblock \bibinfo{journal}{\emph{Quantum}}  \bibinfo{volume}{2}
  (\bibinfo{year}{2018}), \bibinfo{pages}{79}.
\newblock


\bibitem[Rudolph et~al\mbox{.}(2023)]%
        {rudolph2023decomposition}
\bibfield{author}{\bibinfo{person}{Manuel~S Rudolph}, \bibinfo{person}{Jing
  Chen}, \bibinfo{person}{Jacob Miller}, \bibinfo{person}{Atithi Acharya},
  {and} \bibinfo{person}{Alejandro Perdomo-Ortiz}.}
  \bibinfo{year}{2023}\natexlab{}.
\newblock \showarticletitle{Decomposition of matrix product states into shallow
  quantum circuits}.
\newblock \bibinfo{journal}{\emph{Quantum Science and Technology}}
  \bibinfo{volume}{9}, \bibinfo{number}{1} (\bibinfo{year}{2023}),
  \bibinfo{pages}{015012}.
\newblock


\bibitem[Shende et~al\mbox{.}(2006)]%
        {1629135}
\bibfield{author}{\bibinfo{person}{V.V. Shende}, \bibinfo{person}{S.S.
  Bullock}, {and} \bibinfo{person}{I.L. Markov}.}
  \bibinfo{year}{2006}\natexlab{}.
\newblock \showarticletitle{Synthesis of quantum-logic circuits}.
\newblock \bibinfo{journal}{\emph{IEEE Transactions on Computer-Aided Design of
  Integrated Circuits and Systems}} \bibinfo{volume}{25}, \bibinfo{number}{6}
  (\bibinfo{year}{2006}), \bibinfo{pages}{1000--1010}.
\newblock
\urldef\tempurl%
\url{https://doi.org/10.1109/TCAD.2005.855930}
\showDOI{\tempurl}


\bibitem[Shor(1999)]%
        {shor1999polynomial}
\bibfield{author}{\bibinfo{person}{Peter~W Shor}.}
  \bibinfo{year}{1999}\natexlab{}.
\newblock \showarticletitle{Polynomial-time algorithms for prime factorization
  and discrete logarithms on a quantum computer}.
\newblock \bibinfo{journal}{\emph{SIAM review}} \bibinfo{volume}{41},
  \bibinfo{number}{2} (\bibinfo{year}{1999}), \bibinfo{pages}{303--332}.
\newblock


\bibitem[Soklakov and Schack(2006)]%
        {Soklakov2006}
\bibfield{author}{\bibinfo{person}{Andrei~N. Soklakov} {and}
  \bibinfo{person}{Rüdiger Schack}.} \bibinfo{year}{2006}\natexlab{}.
\newblock \showarticletitle{Efficient state preparation for a register of
  quantum bits}.
\newblock \bibinfo{journal}{\emph{Physical Review A}} \bibinfo{volume}{73},
  \bibinfo{number}{1} (\bibinfo{date}{jan} \bibinfo{year}{2006}),
  \bibinfo{pages}{012307}.
\newblock
\urldef\tempurl%
\url{https://doi.org/10.1103/physreva.73.012307}
\showDOI{\tempurl}


\bibitem[Trugenberger(2001)]%
        {Trugenberger2001}
\bibfield{author}{\bibinfo{person}{C.~A. Trugenberger}.}
  \bibinfo{year}{2001}\natexlab{}.
\newblock \showarticletitle{Probabilistic Quantum Memories}.
\newblock \bibinfo{journal}{\emph{Physical Review Letters}}
  \bibinfo{volume}{87}, \bibinfo{number}{6} (\bibinfo{date}{jul}
  \bibinfo{year}{2001}), \bibinfo{pages}{067901}.
\newblock
\urldef\tempurl%
\url{https://doi.org/10.1103/physrevlett.87.067901}
\showDOI{\tempurl}


\bibitem[Ventura and Martinez(1999)]%
        {ventura1999initializing}
\bibfield{author}{\bibinfo{person}{Dan Ventura} {and} \bibinfo{person}{Tony
  Martinez}.} \bibinfo{year}{1999}\natexlab{}.
\newblock \showarticletitle{Initializing the amplitude distribution of a
  quantum state}.
\newblock \bibinfo{journal}{\emph{Foundations of Physics Letters}}
  \bibinfo{volume}{12}, \bibinfo{number}{6} (\bibinfo{year}{1999}),
  \bibinfo{pages}{547--559}.
\newblock


\bibitem[Wille et~al\mbox{.}(2022)]%
        {wille2022decision}
\bibfield{author}{\bibinfo{person}{Robert Wille}, \bibinfo{person}{Stefan
  Hillmich}, {and} \bibinfo{person}{Lukas Burgholzer}.}
  \bibinfo{year}{2022}\natexlab{}.
\newblock \showarticletitle{Decision diagrams for quantum computing}.
\newblock In \bibinfo{booktitle}{\emph{Design automation of quantum
  computers}}. \bibinfo{publisher}{Springer}, \bibinfo{pages}{1--23}.
\newblock


\bibitem[{Wootters} and {Zurek}(1982)]%
        {1982Natur.299..802W}
\bibfield{author}{\bibinfo{person}{W.~K. {Wootters}} {and}
  \bibinfo{person}{W.~H. {Zurek}}.} \bibinfo{year}{1982}\natexlab{}.
\newblock \showarticletitle{{A single quantum cannot be cloned}}.
\newblock \bibinfo{journal}{\emph{Nature}} \bibinfo{volume}{299},
  \bibinfo{number}{5886} (\bibinfo{date}{Oct.} \bibinfo{year}{1982}),
  \bibinfo{pages}{802--803}.
\newblock
\urldef\tempurl%
\url{https://doi.org/10.1038/299802a0}
\showDOI{\tempurl}


\bibitem[Wu et~al\mbox{.}(2021)]%
        {PhysRevLett.127.180501}
\bibfield{author}{\bibinfo{person}{Yulin Wu} {et~al\mbox{.}}}
  \bibinfo{year}{2021}\natexlab{}.
\newblock \showarticletitle{Strong Quantum Computational Advantage Using a
  Superconducting Quantum Processor}.
\newblock \bibinfo{journal}{\emph{Phys. Rev. Lett.}}  \bibinfo{volume}{127}
  (\bibinfo{date}{Oct.} \bibinfo{year}{2021}), \bibinfo{pages}{180501}.
\newblock
Issue 18.
\urldef\tempurl%
\url{https://doi.org/10.1103/PhysRevLett.127.180501}
\showDOI{\tempurl}


\end{thebibliography}

\end{document}